\documentstyle[11pt,aaspp4,psfig]{article}

\received{1998 January 22}
\accepted{1998 February 12}
\begin{document}     
\def\today{\ifcase\month\or
January\or February\or March\or April\or May\or June\or
July\or August\or September\or October\or November\or December\fi
\space\number\day, \number\year}

\newcommand{\squig}{$\sim$}
\newcommand{\squigleq}{\mbox{$^{<}\mskip-10.5mu_\sim$}}
\newcommand{\squiggeq}{\mbox{$^{>}\mskip-10.5mu_\sim$}}
\newcommand{\squiggeqmm}{\mbox{$^{>}\mskip-10.5mu_\sim$}}
\newcommand{\decsec}[2]{$#1\mbox{$''\mskip-7.6mu.\,$}#2$}
\newcommand{\decsecmm}[2]{#1\mbox{$''\mskip-7.6mu.\,$}#2}
\newcommand{\decdeg}[2]{$#1\mbox{$^\circ\mskip-6.6mu.\,$}#2$}
\newcommand{\decdegmm}[2]{#1\mbox{$^\circ\mskip-6.6mu.\,$}#2}
\newcommand{\decsectim}[2]{$#1\mbox{$^{\rm s}\mskip-6.3mu.\,$}#2$}
\newcommand{\decmin}[2]{$#1\mbox{$'\mskip-5.6mu.$}#2$}
\newcommand{\asecbyasec}[2]{#1$''\times$#2$''$}
\newcommand{\aminbyamin}[2]{#1$'\times$#2$'$}

\title{Complex Velocity Fields in the Shell of T Pyxidis}
\author{Bruce Margon and Eric W. Deutsch}
\affil{Department of Astronomy, 
       University of Washington, Box 351580,
       Seattle, WA 98195-1580\\
       margon@astro.washington.edu; deutsch@astro.washington.edu}

\begin{center}
Accepted for publication in The Astrophysical Journal Letters\\
{\it received 1998 January 22; accepted 1998 February 12}
\end{center}

\begin{abstract}

We present spatially-resolved, moderate-resolution spectrophotometry of 
the recurrent nova T~Pyx and a portion of the surrounding 
shell. The spectrum extracted from a strip of width $10''$ centered on 
the star shows well-known, strong emission lines typical of old novae, plus a 
prominent, unfamiliar emission line at $\lambda$6590.
This line, and a weaker companion at $\lambda$6540 which we also detect, have 
been previously reported by Shahbaz et al., and attributed to 
Doppler-shifted H$\alpha$ emission from a collimated jet emerging from T~Pyx.
We demonstrate that these lines are instead due to [N\,II] 
$\lambda\lambda$6548, 6584 from a complex velocity field in the surrounding 
nebula. The comments of past workers concerning
the great strength of He\,II
$\lambda$4686 in T~Pyx itself are also reiterated.

\end{abstract}


\clearpage
\section{INTRODUCTION}

The recurrent nova T~Pyxidis was discovered by Leavitt (1914); a history of the 
object, and some recent observations, may be found in Webbink et al. (1987) 
and Shara et al. (1989). A compact ($\sim10''$) nebula surrounding the star
is described by Duerbeck \& Seitter (1979) and Williams (1982), and a larger,
fainter shell is observed by Shara et al. (1989). {\it Hubble Space
Telescope} imagery of the T~Pyx nebula (Shara et al. 1997) reveals an
exceptionally complex, clumpy structure on subarcsecond spatial scales, and
the brightness of at least some of these knots varies on timescales of months.
As pointed out by these authors, the {\it HST} data vividly demonstrate that 
``shell" is a quite misleading term for the extended structure near T~Pyx, 
which in fact consists of literally thousands of discrete lumps, most 
bright in [N\,II] emission.

Recently an interesting spectrum of T~Pyx has been presented by Shahbaz et al. 
(1997).  They call attention to a strong, unfamiliar emission line at 
$\lambda6593$, and a weaker feature at $\lambda6539$. They interpret these 
features as Doppler-shifted H$\alpha$ lines from a collimated jet emerging 
from T~Pyx, at velocities of +1400 and $-1100$~km~s$^{-1}$. As there is no {\it
a priori} reason to expect these particular velocities, and no second spectral
feature to confirm them, care must obviously be taken when accepting this
interpretation; virtually any unidentified emission line, regardless of
wavelength, can be attributed to such a model. Nonetheless this
interpretation if correct is very exciting: as stressed by Shahbaz et al.
(1997), this would make T~Pyx the first short-period cataclysmic variable
with a jet. Livio (1998), who comments that ``jet lines have now been
observed unambiguously in the recurrent nova T~Pyx," stresses that there
would be profound implications on models of jet formation.

In an effort to clarify this unprecedented interpretation, we obtained further
spectra of T~Pyx and the surrounding nebula. Although we verify the existence 
of the unusual emission lines, the features unfortunately prove to have a less
exotic origin than that suggested by Shahbaz et al.

\section{OBSERVATIONS AND INTERPRETATION}

On UT 1997 December~6, we obtained spectrophotometry of T~Pyx with the 3.5-m 
telescope of the Astrophysical Research Consortium, located at Apache Point, 
NM, using the Double Imaging Spectrograph (DIS) in its high-resolution mode. 
In this configuration, DIS simultaneously provides two spectra, each of about
1000~\AA\ coverage in disjoint wavelength ranges, split by a dichroic mirror
onto two CCDs. A slit measuring \decsec{1}{5} $\times\ 5'$ was employed; the
measured FWHM of comparison arc lines indicates that a spectral resolution of
$\sim2.4$~\AA\ was achieved. The slit was oriented E/W, passing through the 
star as well as a series of the bright knots in the nova shell located within
a few arcsec of the central 
object (Shara et al. 1997).

At $\delta=-32^\circ$, T~Pyx is not well-located for observation from Apache 
Point, so observations were confined to two hours centered on the meridian. A 
total of 6600~s of integration in 6 separate exposures was obtained, and the 
data summed together for analysis. Approximate flux calibration was achieved
via observation of spectrophotometric standard stars from Massey et al.
(1988) and Massey \& Gronwall (1990). However at these large airmasses, the 
combination of differential refraction and uncertain light losses at the slit
are such that these absolute fluxes cannot be regarded as more accurate than
$\pm0.5$~mag.

The two resulting spectra appear in Figure~1. Strong Balmer and He\,I emission
are evident, as is typical in many old novae, as well as the prominent
C\,III/N\,III $\lambda\lambda4640, 4650$ blend and especially He\,II
$\lambda$4686. The spectrum is similar to that displayed by Williams (1983)
and described by Duerbeck \& Seitter (1987). We reiterate the remarks by
Williams (1983) and
Williams (1989) concerning the great strength of He\,II $\lambda$4686, which
equals or exceeds H$\beta$ in intensity. As this latter circumstance is
usually seen only in magnetic cataclysmic variables, the polarization
properties of T~Pyx are clearly of great interest and should be determined.
Because the extraction width of the data in Figure~1 is $10''$, the stellar
Balmer and He\,II emission may also contain a component due to the
surrounding nebula. Low resolution spectral scans of the north portion of the
nebula, uncontaminated by the star but spatially disjoint from the region
observed in this work, by Williams (1982) show a quite small value of
$\lambda$4686/H$\beta$, hinting but certainly not proving that the unusual
ratio seen in the star is truly a valid measure.

Immediately evident in the red spectrum is a prominent, resolved emission line
centered at $\lambda$6590, which includes the unusual feature discovered by 
Shahbaz et al. (1997), and termed ``S$^+$" by those authors. It is not
obviously visible in the lower-resolution data of Williams (1983). In
Figure~2 we present intensity contour maps of portions of the spectra
displayed in Figure~1, but with the stellar continuum subtracted. Here we also
include the spatially-resolved data in the region $\pm10''$ from T~Pyx,
twice the areal coverage of Figure~1. In these maps, the abscissa is
wavelength, and the ordinate is the spatial direction
E/W of the central star. Even with the modest pixel scale of DIS 
(\decsec{0}{61}~pix$^{-1}$ in the red, and \decsec{1}{1}~pix$^{-1}$ in the 
blue), it is clear that substantial flux is detected in our data from the 
nebula, spatially resolved from the central star.

Although extensive dispersive nebular spectroscopy of T~Pyx has not to our 
knowledge been presented, it is evident from the work of Williams (1982) and 
Shara et al. (1989, 1997) that [N\,II] $\lambda$6584 is very prominent in the
knots of the nebula. In agreement with this past work, examination of the 
contours in Figure~2 readily shows that we have also detected [N\,II] 
$\lambda$6584, together with associated [N\,II] $\lambda$6548, H$\alpha$, and 
[O\,III] $\lambda\lambda$ 4959, 5007 emission, at a number of separable, 
discrete velocities (presumably each corresponding to one or a small number of
bright nebular knots) within a few arcseconds of T~Pyx.

Interpretation of these data in a two-dimensional plane such as that of 
Figure~2 can be confusing, as distinct spatial regions can overlap in 
projection, and velocity differences of $\sim10^3$~km~s$^{-1}$ can result in
the wavelength blending of H$\alpha$ and the two red [N\,II] lines in one
knot with different components of this same triple of lines from other,
discrete velocity systems of another knot. To aid in identification of the
strongest, most prominent velocity systems, we place on Figure~2 colored bars
at the location of [O\,III] $\lambda\lambda$ 4959, 5007, H$\alpha$, and
[N\,II] $\lambda\lambda$6548, 6584, using a distinct color for each of five
identified velocity systems which are prominent. Further velocity structure
is evident, but more sophisticated representations of the results are
probably not warranted by the relatively crude spectral and spatial
resolution of our data. In any case, this analysis suffices to illustrate our
most important conclusions. While we have included the appropriate
heliocentric correction of $\sim20$~km~s$^{-1}$, the absolute velocity
calibration of the spectrograph is uncertain at the level of
$\sim10$~km~s$^{-1}$; relative velocities should be more reliable. A slight
additional complication is that due to differential refraction at this high
airmass, the red and blue sides of the spectrograph are observing slightly
different, although spatially adjacent, regions of the nebula separated by
$\sim1''$ (Filippenko 1982); this small effect has no impact on the
conclusions reached here.

We first call attention to the ``orange" system, at $v\sim+250$~km~s$^{-1}$.
This is very clearly nebular gas, detected simultaneously in five separate
emission lines (four of them forbidden) at a consistent velocity, and
colocated $3-6''$ W of the star. Now consider the nearby ``red" system, which
lies essentially coincident with the stellar image in the E/W direction.
Figure~2 identifies the emission at $\lambda$6595 as [N\,II] $\lambda$6584 at
$v\sim+490$~km~s$^{-1}$. For this velocity, [N\,II] $\lambda$6548 and
H$\alpha$ are blended with the ``purple" and ``blue" systems, blue-shifted gas
which probably falls only in projection, rather than spatial coincidence, 
on our image. However, note that the ``red" system does appear as expected in 
[O\,III] $\lambda$4959 (and cannot be expected to appear in [O\,III] 
$\lambda$5007 due to a precise overlap with the strong He\,I $\lambda$5015 from 
the star). Furthermore, although it may be a projection effect, the most 
straightforward interpretation of the orange and red systems is that as they 
are reasonably adjacent in both velocity and E/W location, they are in fact 
physically adjacent in the nebula, and both are emission systems from knots in
the shell.

Figure~2 shows clearly that in our data the [N\,II] $\lambda$6584 emission
from the orange and red systems combine to form the prominent,
spectrally-resolved ``S$^+$" emission line at $\lambda$6590. Although Shahbaz
et al. (1997) do not provide details of the shape and size of the slit or
aperture used to obtain their data, and in any case {\it post facto}
reconstruction of the precise pointing location and seeing is probably
not possible, we suggest that the ``S$^+$" described by these authors is also
red-shifted [N\,II] $\lambda$6584 from knots in the nebula, either physically
close to, or seen in projection close to, T~Pyx. Indeed, inspection of
Figure~2 shows that a narrow, on-axis aperture would produce emission
centered at $\lambda6593$, exactly as they have reported.

We believe that a similar explanation applies to the $\lambda$6539 
``S$^-$" emission line reported by 
Shahbaz et al., and again attributed by them to H$\alpha$ emission from 
a collimated jet emitted by the star.  We also detect this line, which we 
denote as the ``violet" system in Figure~2. We suggest that the ``S$^-$" line
is in fact [N\,II] $\lambda$6548 at $v= -440$~km~s$^{-1}$. As this system is
spatially projected very close to the star, other lines at the same velocity
are difficult to discern, but are indeed present. For example, the
corresponding [N\,II] $\lambda$6584 and [O\,III] $\lambda$5007 lines in the
violet system do appear as intensity peaks at the correct wavelengths,
superposed on the stellar continuum, and are visible in Figure~2 as excess 
contours.

Figure~2 illustrates that even given our meager data at one slit position 
angle, the velocity structure of the nebula is clearly complex. Shara et al. 
(1997) have demonstrated that the very small scale spatial structure of the 
knots is exceptionally intricate, and that knot intensities can fade and grow 
on a timescale of months.  Taken together this implies that ground-based 
spectral observations are not ideal to probe the velocity field, and that 
great caution must be employed when comparing data obtained even with the same
equipment at different times, much less by different observers with different 
facilities. One can predict in advance that very minor differences in 
positioning of an entrance aperture, when convolved with the great spatial 
complexity of the shell, variable seeing, and 
the intrinsic flux variations of the object, will conspire to at least 
slightly change the observed profile, intensity, and mean wavelength of the 
$\lambda$6590 emission line.

\section{CONCLUSION}

We have presented spatially-resolved spectrophotometry of T~Pyx and portions 
of the surrounding nebula. Recent {\it HST} images have vividly stressed how
spatially complex the extended structure is. From our data, it is also clear
that the shell is spectrally complex. We believe that there is little
evidence that the
prominent $\lambda$6590 emission line and its weaker companion at
$\lambda$6540 are due to H$\alpha$ from a collimated jet ejected by the star
(Shahbaz et al. 1997). Rather, we argue, based on the detection of multiple
emission lines from different species at consistent velocities, that these
lines are instead [N\,II] from a few, or possible even many, discrete knots.
The large He\,II
$\lambda$4686/H$\beta$ ratio, probably due to T~Pyx itself, also bears
further scrutiny.

\acknowledgments

We appreciate several conversations with G.~Wallerstein and H.~M.~Schmid
on the spectrum of T~Pyx.


\clearpage

\begin{figure}
\plotone{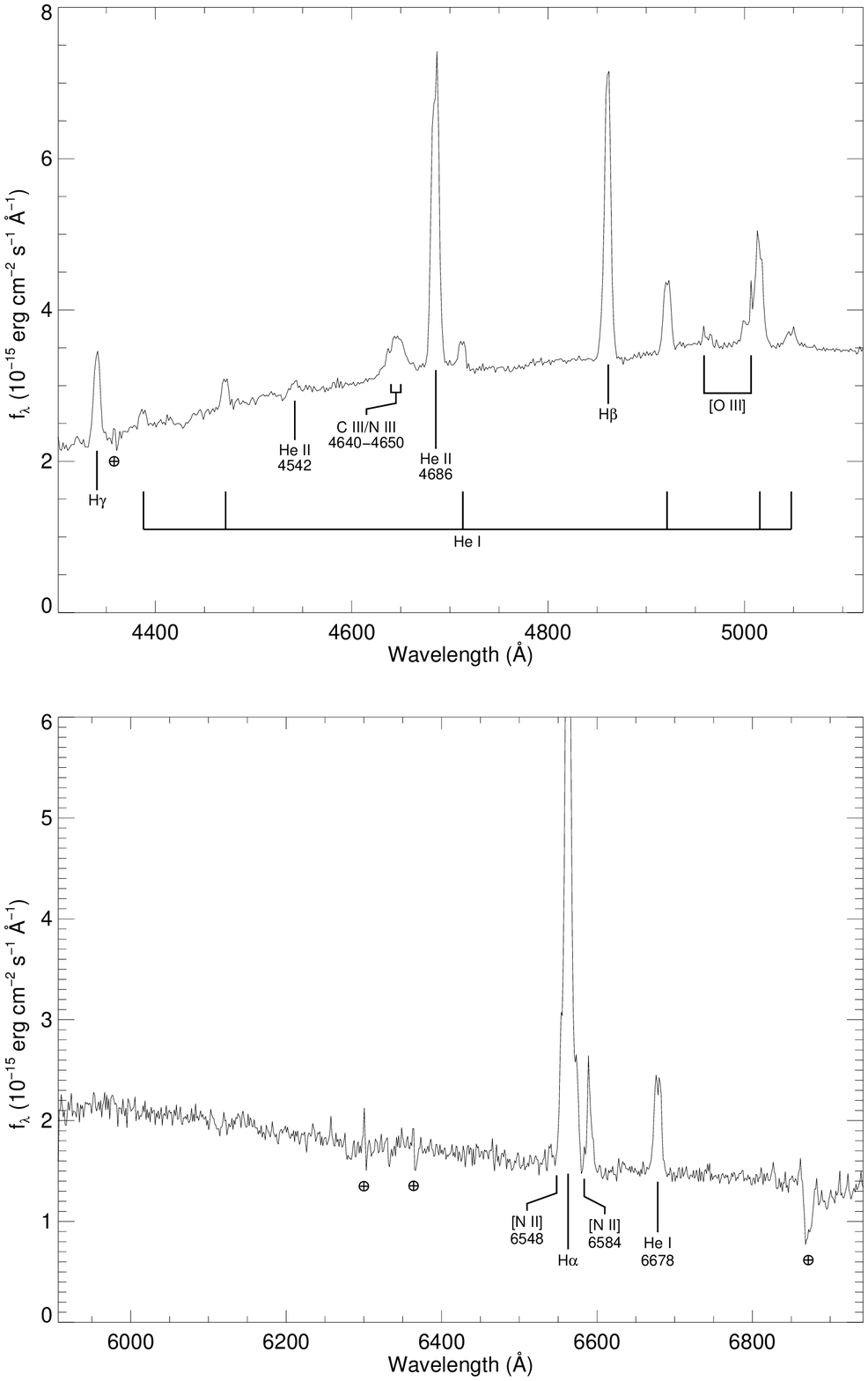}
\caption{
The spectrum of T~Pyx and the adjacent $\pm5''$ to the East/West.
The very strong H$\alpha$ emission 
line has been truncated for convenience in scaling; the observed peak 
intensity is $8\times10^{-15}$. The continuum slope, especially in the blue, 
is unreliable due to the high airmass of the observations; indeed, the object 
is known to be very UV-excess. The same factor limits the precision of the 
absolute fluxes to of order $\pm0.5$~mag. Note the apparent strength of the 
anomalous $\lambda$6590 emission feature, discussed in the text.
}
\end{figure}

\begin{figure}
\hskip0.9in\psfig{figure=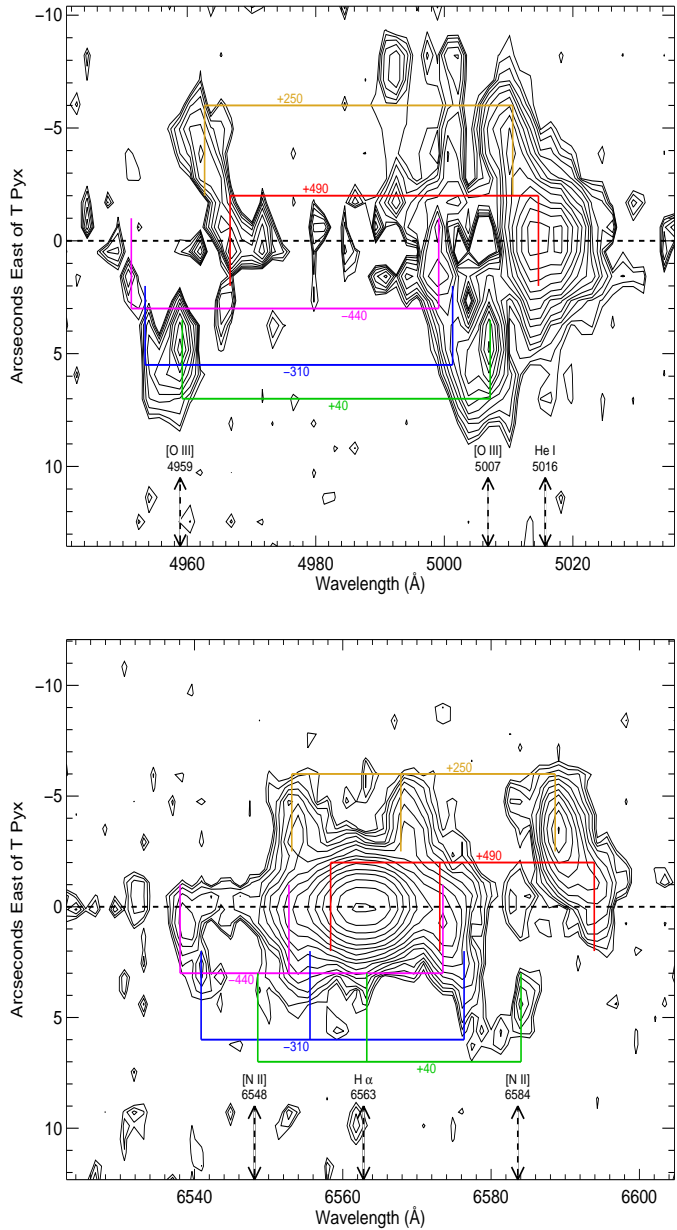,height=7in,width=4.575in}
\caption{
Intensity contours of spatially-resolved, continuum-subtracted spectra of
T~Pyx and the surrounding region. These data are identical to those of
Figure~1, but in addition include the region $\pm10''$ East and West of the
star. Each contour is $2\times$ the intensity of the previous one.
Each colored bar identifies a system of nebular gas at one consistent 
velocity, calculated from the detection of two or more emissions lines from the
group of [N\,II] $\lambda\lambda$6548, 6584, H$\alpha$, and 
[O\,III] $\lambda\lambda$ 4959, 5007; the reddest colors have the largest 
recessional velocities, and the bluest, the largest approach. The absolute,
heliocentric-corrected
velocity values suggested for each system may have uncertainties as large as 
$\pm10$~km~s$^{-1}$, but relative velocities should be substantially more 
accurate. For reference, 
the broken lines show the rest wavelengths of these emission features, as
well as that of the He\,I $\lambda5016$ line, which is detected strongly in the
stellar spectrum. We suggest that the prominent, anomalous $\lambda6590$ 
emission line (Figure~1) is formed by the superposition of red-shifted [N\,II]
$\lambda6584$ from the red and orange systems, seen very clearly in this 
Figure.
}
\end{figure}

\end{document}